\documentclass[%
preprint,
showpacs,preprintnumbers,
 amsmath,amssymb,
 aps,
]{revtex4-1}

\usepackage{graphicx}
\usepackage{dcolumn}
\usepackage{bm}
\usepackage[dvipdfm]{hyperref}

\begin{document}
\title{Quantum Szilard Engines with arbitrary spin}
\author{Zekun Zhuang}
\author{Shi-Dong Liang}
\altaffiliation{Corresponding author: Email: stslsd@mail.sysu.edu.cn}
\affiliation{State Key Laboratory of Optoelectronic Material and Technology, and
Guangdong Province Key Laboratory of Display Material and Technology,
School of Physics and Engineering, Sun Yat-Sen University, Guangzhou, 510275,
People's Republic of China}
\date{\today }

\begin{abstract}
The quantum Szilard engine (QSZE) is a conceptual quantum engine for understanding the fundamental physics of quantum thermodynamics and information physics. We generalize QSZE to arbitrary spin case, i.e. spin QSZE (SQSZE) and systematically study the basic physical properties of both fermion and boson SQSZEs in low temperature approximation. We give the analytic formulation of the total work. For the fermion SQSZE, the work might be absorbed from the environment, and the change rate of the work with temperature exhibits periodicity and even-odd oscillation, which is a generalization of spinless QSZE.
It is interesting that the average absorbed work oscillates regularly and periodically in large-number limit, which implies that the average absorbed work in fermion SQSZE
is neither an intensive quantity nor an extensive quantity. The phase diagrams of both fermion and boson SQSZEs gives the SQSZE doing positive or negative work in the parameter space of the temperature and the particle number of system, but they have different behaviors because the spin degrees of fermion and boson play different roles in their configuration states and  corresponding statistical properties. The critical temperature of phase transition depends sensitively on the particle number.
By using the Landauer's erasure principle, we give the erasure work in a thermodynamic cycle and define a new efficiency (we refer it as information-work efficiency) to measure the engine's ability of utilizing information to extract work. We also give the conditions under which the maximum extracted work and highest information-work efficiencies for fermion and boson SQSZEs can be achieved.

\end{abstract}

\maketitle



\section{Introduction}
\begin{figure}
\includegraphics[width=5 in]{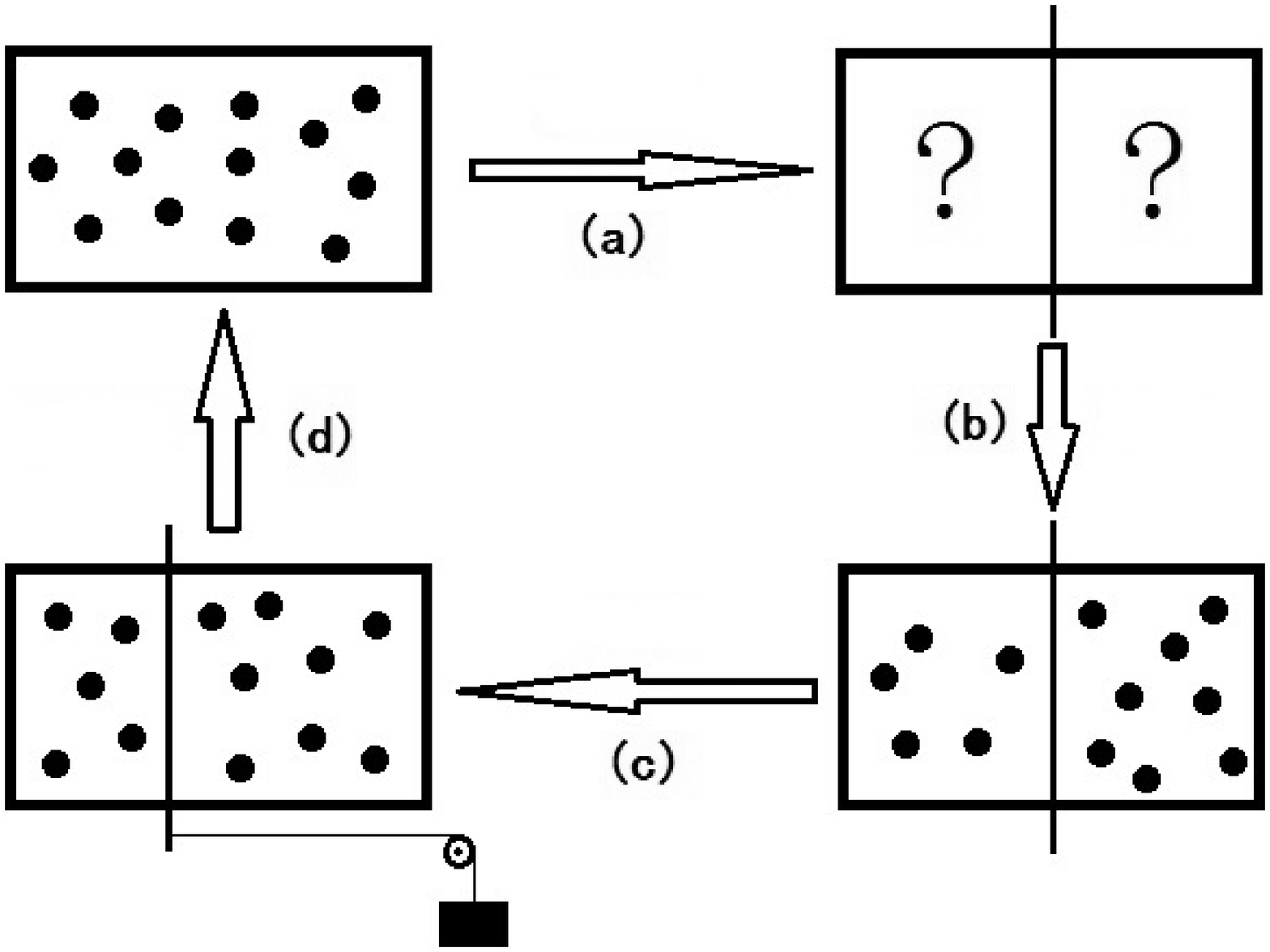}
\caption{\label{fig1}
The working cycle of the classical multi-particle Szilard engine. (a) A wall is isothermally inserted into the box.
(b) Then a measurement is performed to know the number of particles of both sides.
(c) The system does work by isothermally moving the wall to the equilibrium position.
(d) The wall is removed isothermally and the system is restored to the initial state.
}
\end{figure}

The Szilard engine (SZE) is a single molecule engine designed for solving
the Maxwell's demon paradox\cite{szilard}. As is shown in
Fig. \ref{fig1},
$N$ ideal identical particles are prepared in a one-dimensional box
of length $L$, and the system is always in contact with a heat bath of
temperature $T$. Then an adiabatic wall is isothermally inserted in the box
at position $l$, after which a measurement is performed to find $m$
particles in the left room. As long as the number of particles on each side
are unequal, the wall would move to equilibrium position $l_{eq}^{m}$ and perform work
through isothermal expansion. Finally, the wall is removed and the whole
thermodynamic cycle is completed. All the processes of the cycle are
isothermal quasistatic processes.

Szilard analyzed this engine and concluded that the entropy produced during the measurement saves the second law of thermodynamics\cite{szilard}.
Bennett and Landauer
argued that erasuring information (demon's memory) generates entropy $%
k_{B}\ln 2$, where $k_{B}$ is the Boltzmann constant, which compensates the
entropy decreasing in the thermodynamic cycle\cite{bennett,landauer}. Now it is widely accepted that the total entropy generated during the
measurement process and the erasure of demon's memory compensates the entropy
decrease of the engine \cite{sagawa}.This reveals the relationship between physics and information,
which leads to much interest in information in thermodynamic cycle and
quantum heat engine \cite{arnaud,eitan,geusic}.

Recently, Kim \textit{et al.} gave a formula of the total work in a thermodynamic
cycle for the quantum Szilard engine (QSZE), which is related to the
relative entropy in the classical limit \cite{kim}. In the QSZE with two
indistinguishable identical particles in the infinite potential well they
found that indistinguishability plays an important role in the total work in a
thermodynamic cycle \cite{kim}.  Lu \textit{et al.} generalized this QSZE to contain arbitrary
number of identical particles \cite{lu}. They found that in high-temperature
limit, the works in a cycle are equal for boson and fermion QSZE, which can
be understood by the transition from quantum to classical statistics when
temperature increases. However, in low-temperature limit, the physical
scenario between fermion and boson QSZE is quite different. The fermion QSZE
shows the parity effect, namely, the QSZE can extract work in a
thermodynamic cycle for odd number of fermions, but cannot extract work for
even number of fermions. For boson QSZEs, there exists a phase transition at
a critical temperature for extracted work to be positive or negative.\cite{lu} These
effects of QSZE imply the significant role of spin nature in
low-temperature limit. However, QSZE studied by Kim's and Lu's group ignore
the spin degree of freedom of particles. In fact, it can be seen later that the spin would result in the degeneracy of energy levels, which would alter the configuration states of the system in low-temperature limit. So it should be important and
interesting for the spin effect in QSZE.

In this paper, we will generalize QSZE to include arbitrary spin and
arbitrary number of fermion and boson working substance. We will give the
analytic expressions of the extracting work in a thermodynamic cycle in low-temperature approximation, in
terms of the number, spin and temperature. We will also show a detail
analysis of the thermodynamic properties in a thermodynamic cycle with the
particle number, temperature and spin of fermion and boson, including the
total work, absorbed work and phase diagram of spin QSZE (SQSZE). We introduce a new
efficiency and give the condition of maximizing work of SQSZE.

\section{Model of Spin Quantum Szilard Engines}

We consider a spin QSZE (SQSZE) that contains $N$ fermions or bosons with spin $s$ as the working substance working in a one-dimensional infinite potential well. The thermodynamic cycle of SQSZE consists of four processes,
namely, insertion, measurement, expansion and removal for four distinct
states. The total work in a thermodynamic cycle can be expressed as \cite{kim}

\begin{equation}
W_{tot}=-k_{B}T\sum_{m=0}^{N}f_{m}\ln \left( \frac{f_{m}}{%
f_{m}^{\ast }}\right)
\label{workexp}
\end{equation}
where $f_{m}=\frac{Z_{m}(\ell)}{Z(\ell)}$ is the probability for having $m$
particles on the left side after measurement and $Z(\ell)=$ $%
\sum_{m=0}^{N}Z_{m}(\ell)$ represents the partition function for the whole
system before measurement. $Z_{m}(\ell)$ is the partition function for the case
in which the wall is at the position $\ell$ and $m$ particles are on the left and $%
N-m$ on the right. $f_{m}^{\ast }$ is defined as $f_{m}^{\ast }=\frac{%
Z_{m}(\ell_{eq}^{m})}{Z(\ell_{eq}^{m})}$, where $Z(\ell_{eq}^{m})=$ $%
\sum_{n=0}^{N}Z_{n}(\ell_{eq}^{m})$ and $Z_{n}(\ell_{eq}^{m})$ represents the
partition function for the case that the wall is at the equilibrium position $%
\ell_{eq}^{m}$ and $n$ particles are on the left. $k_{B}$ is the Boltzmann
constant and $T$ is the temperature of the heat bath.

The partition function can be written as $Z=\sum_{s}e^{-E_{s}/k_{B}T}$ where the sum runs all states of the system. However, in low-temperature limit, i.e. $T\rightarrow 0$, it can be assumed that all particles occupy the possible lowest-energy states and  all systems in the ensemble are in the ground state. Namely, the high energy levels and excited states of the system can be ignored in the low-temperature limit. The partition function can be rewritten as
\begin{equation}
Z=g_{G}e^{-E_{g}/k_{B}T}
\end{equation}
where $g_{G}$ is the degeneracy of ground state of the system and $E_{g}$ is the ground-state energy of the system.

Without losing generality, the wall is assumed to be inserted at the
position $\ell=\frac{L}{2}$, where $L$ is the well width, and initially
particles are working in the energy eigenstates, $E_{n}(\ell)=\frac{n^{2}\pi
^{2}\hbar ^{2}}{2M{\ell}^{2}}$, where $n=1,2...$labels eigenenergy levels.

\subsection{Spin fermion case}
Let us consider $N$ spin-$s$ fermions as the working particles. For convenience, we define variable $u=s+1/2$ so that $u$ is an integer. Suppose that the total particle number is $N=4un+k$, where $0\leq k<4u$. The fermions occupy the energy levels following the Pauli's principle. Namely, each energy level of infinite quantum well can be occupied with at most $2s+1$ particles.
In general, there are two cases for $k$, $0\leq k<2u$ and $2u\leq k<4u$. They have different configuration states and corresponding different partition functions. We will discuss these two cases respectively in the following sections.

\begin{figure}
\includegraphics[width=5 in]{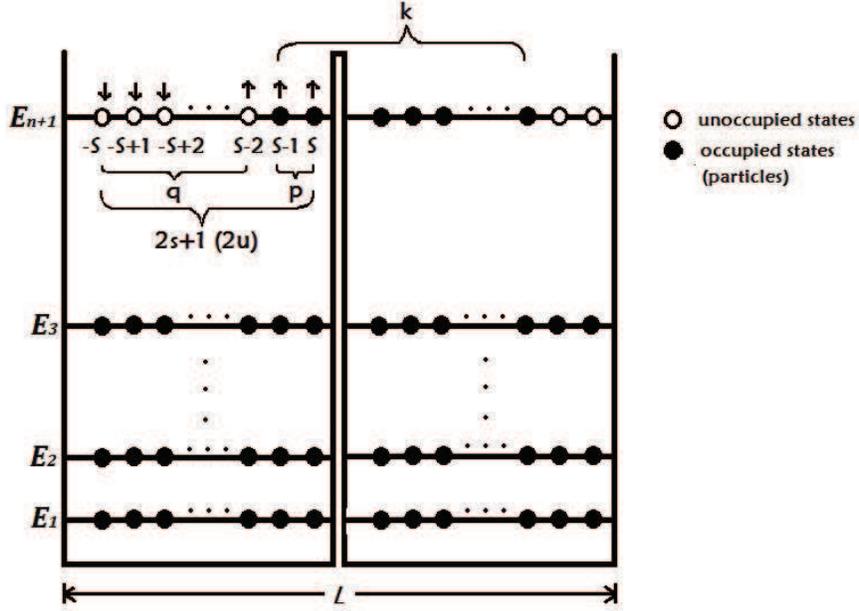}
\caption{\label{fig2}
The energy state of the fermion SQSZE when the wall is inserted at position $ L/2 $ in the low-temperature limit
}
\end{figure}

{\bf Case 1}: $0\leq k<2u$,
suppose that $m$ particles are on the left side of the well after measurement. We can conclude that in the low-temperature limit it is impossible for $0\leq m<2un$ or $2un+k<m\leq4un+k$, since $Z_{m}(L/2)\ll Z(L/2)$ and $f_{m}\rightarrow0$. So for the case that the wall is at the initial position $\ell=L/2$, there are $2un$ particles occupying the energy levels from $1,2,...$ to $n$th states on the left and right sides of the well respectively, and $k$ particles occupy the $(n+1)$th state on the left or right side of the well, as is shown in Fig. \ref{fig2}.
It implies $p=m-2un$ particles occupying the $(n+1)$th energy level of the left well  where $p$ is within $0\leq p\leq k$.
By using the combinatorics to calculate the degeneracy of the ground state (see Appendix A), and the partition function, we can obtain

\begin{equation}
f_{m}=\left\{
\begin{array}{lll}
\frac{C_{2u}^{p}C_{2u}^{k-p}}{C_{4u}^{k}} & , & 0\leq p\leq k \\
0 & , & otherwise%
\end{array}%
\right.
\label{Ffm-1}
\end{equation}
in the low-temperature limit, where $C_a^b$ represents combinatorial number.

When the wall moves to the position $l_{eq}^{m}$, there are $2un$ particles occupying the energy levels from $1,2,...$ to $n$th states on the left and right sides of the well respectively. For the case that $m$ particles are limited on the left side, $p$ particles occupy the $(n+1)$th state on the left side of the well and $k-p$ particles occupy the $(n+1)$th state on the right side of the well. For the case that particles are not limited on one side due to the tunnelling effect \cite{kim}, remain $k$ particles occupy the $(n+1)$th state on one side of the well, in which the energy of the $(n+1)$th energy level is lower than that of the other side. So for $k$ is odd, we get
\begin{subequations}
\label{Ffm*1whole}
\begin{equation}
f_{m}^{\ast }=\frac{C_{2u}^{p}C_{2u}^{k-p}}{C_{2u}^{k}}e^{-\mu \beta \Delta
E_{n+1}^{m}}
\label{Ffm*1-odd}
\end{equation}
and for $k$ is even,
\begin{equation}
f_{m}^{\ast }=\left\{
\begin{array}{lll}
\frac{C_{2u}^{p}C_{2u}^{k-p}}{C_{2u}^{k}}e^{-\mu \beta \Delta E_{n+1}^{m}} &
, & p\neq \frac{k}{2} \\
\frac{\left( C_{2u}^{k/2}\right) ^{2}}{C_{4u}^{k}} & , & p=\frac{k}{2}%
\end{array}%
\right.
\label{Ffm*1-even}
\end{equation}
\end{subequations}
where $\Delta E_{n+1}^{m}=\left\vert E_{n+1}\left( \ell_{eq}^{m}\right)
-E_{n+1}\left( L-\ell_{eq}^{m}\right) \right\vert $, $\mu =\min \left\{
p,k-p\right\}$ denotes the smaller of the number of particles on the $(n+1)$th level of both sides.

In particular, when the wall reaches the end of the well, $n=0$ and $p=0$ or $k$,
$f_{m}^{\ast }=1$.

By combining the Eqs.(\ref{workexp})(\ref{Ffm-1}) and (\ref{Ffm*1whole}) we can obtain the total work in a thermodynamic cycle of SQSZE,

\begin{equation}
W_{tot}=\left\{
\begin{array}{lll}
k_{B}T\ln \left( \frac{C_{4u}^{k}}{C_{2u}^{k}}\right) -2\sum_{p=1}^{%
\frac{k-1}{2}}p\frac{C_{2u}^{p}C_{2u}^{k-p}}{C_{4u}^{k}}\Delta
E_{n+1}^{2un+p} & \text{for} & k=odd \\
k_{B}T\left[ 1-\frac{\left( C_{2u}^{k/2}\right) ^{2}}{C_{4u}^{k}}\right] \ln
\left( \frac{C_{4u}^{k}}{C_{2u}^{k}}\right) -2\sum_{p=1}^{\frac{k}{2}%
-1}p\frac{C_{2u}^{p}C_{2u}^{k-p}}{C_{4u}^{k}}\Delta E_{n+1}^{2un+p} & \text{%
for} & k=even%
\end{array}%
\right.
\end{equation}
in which we agree that the sum equals to zero if its upper limit is smaller than its lower limit and we obey this agreement in this paper.

{\bf Case 2}: $2u\leq k<4u$,
in this case, notice that there are more than $2u$ particles occupying the $(n+1)$th energy level in the left or right well, which means that particles are more than unoccupied states, we use the hole representation of the state to calculate the partition function for convenience. There are $q=2u(n+1)-m$ unoccupied states on the $(n+1)$th energy level of the left well, where $q$ is within $0\leq q\leq 4u-k$.
By the similar combinatorics, (see Appendix A), we obtain

\begin{equation}
f_{m}=\left\{
\begin{array}{lll}
\frac{C_{2u}^{q}C_{2u}^{4u-k-q}}{C_{4u}^{4u-k}} & , & 0\leq q\leq 4u-k \\
0 & , & otherwise%
\end{array}%
\right.  \label{Ffm-2}
\end{equation}
In the same way, we get
\begin{subequations}
\label{Ffm*2whole}
\begin{equation}
f_{m}^{\ast }=\frac{C_{2u}^{q}C_{2u}^{4u-k-q}}{C_{2u}^{4u-k}}e^{-\lambda
\beta \Delta E_{n+1}^{m}} \label{Ffm*2-odd}
\end{equation}
for odd $k$ and
\begin{equation}
f_{m}^{\ast }=\left\{
\begin{array}{lll}
\frac{C_{2u}^{q}C_{2u}^{4u-k-q}}{C_{2u}^{4u-k}}e^{-\lambda \beta \Delta
E_{n+1}^{m}} & , & q\neq \frac{4u-k}{2} \\
\frac{\left( C_{2u}^{(4u-k)/2}\right) ^{2}}{C_{4u}^{4u-k}} & , & q=\frac{4u-k}{2}
\end{array}%
\right. \label{Ffm*2-even}
\end{equation}
\end{subequations}
for even $k$,
where $\Delta E_{n+1}^{m}=\left\vert E_{n+1}\left( \ell_{eq}^{m}\right)
-E_{n+1}\left( L-\ell_{eq}^{m}\right) \right\vert $, $\lambda =\min \left\{
4u-q-k,q\right\} $ denotes the smaller of the number of unoccupied states on
the $(n+1)$th energy level of both sides.

By the same way, the total work in a thermodynamic cycle of SQSZE can be expressed as

\begin{equation}
W_{tot}=\left\{
\begin{array}{lll}
k_{B}T\ln \left( \frac{C_{4u}^{4u-k}}{C_{2u}^{4u-k}}\right)
-2\sum_{q=1}^{\frac{4u-k-1}{2}}q\frac{C_{2u}^{q}C_{2u}^{4u-k-q}}{%
C_{4u}^{4u-k}}\Delta E_{n+1}^{2u(n+1)-q} & \text{for} & k=odd \\
k_{B}T\left[ 1-\frac{\left( C_{2u}^{(4u-k)/2}\right) ^{2}}{C_{4u}^{4u-k}}%
\right] \ln \left( \frac{C_{4u}^{4u-k}}{C_{2u}^{4u-k}}\right)
-2\sum_{q=1}^{\frac{4u-k}{2}-1}q\frac{C_{2u}^{q}C_{2u}^{4u-k-q}}{%
C_{4u}^{4u-k}}\Delta E_{n+1}^{2u(n+1)-q} & \text{for} & k=even%
\end{array}%
\right.
\end{equation}

Notice that the total works have the same form for both cases, we can rewrite them as
\begin{equation}
W_{tot}=D_{F}k_{B}T-W_{0F} \label{Fworkexp}
\end{equation}
where $D_{F}$ is proportional to the change rate of total work as temperature increases and $W_{0F}$ represents the absorbed work of fermion SQSZE at zero temperature.

For $0\leq k<2u$,
\begin{equation}
D_{F}=\frac{1}{k_{B}}\left( \frac{dW_{tot}}{dT}\right)
=\left\{
\begin{array}{lll}
\ln \left( \frac{C_{4u}^{k}}{C_{2u}^{k}}\right)
& \text{for} & k=odd \\
\left[ 1-\frac{\left( C_{2u}^{k/2}\right) ^{2}}{C_{4u}^{k}}%
\right] \ln \left( \frac{C_{4u}^{k}}{C_{2u}^{k}}\right)
& \text{for} & k=even%
\end{array}%
\right.
\end{equation}
and for $2u\leq k<4u$,

\begin{equation}
D_{F}=\frac{1}{k_{B}}\left( \frac{dW_{tot}}{dT}\right)
=\left\{
\begin{array}{lll}
\ln \left( \frac{C_{4u}^{4u-k}}{C_{2u}^{4u-k}}\right)
& \text{for} & k=odd \\
\left[ 1-\frac{\left( C_{2u}^{(4u-k)/2}\right) ^{2}}{C_{4u}^{4u-k}}%
\right] \ln \left( \frac{C_{4u}^{4u-k}}{C_{2u}^{4u-k}}\right)
& \text{for} & k=even%
\end{array}%
\right.
\end{equation}

\subsection{Spin boson case}
For boson SQSZE, suppose that $N$ spin-$s$ bosons are working in SQSZE and $m$ of them are in the left well after measurement. Each particle has $2s+1$ different spin states, which leads to $2s+1$ degenerate degree of freedom on each energy level. By the combinatorics (see appendix B),  in the low temperature approximation, we can obtain

\begin{equation}
f_{m}=\frac{C_{m+2s}^{2s}C_{N-m+2s}^{2s}}{C_{N+4s+1}^{4s+1}}
\label{Bfm}
\end{equation}
Similarly, we get
\begin{subequations}
\label{Bfm*whole}
\begin{equation}
f_{m}^{\ast }=\left\{
\begin{array}{lll}
\frac{C_{m+2s}^{2s}C_{N-m+2s}^{2s}}{C_{N+2s}^{N}}e^{-\mu \beta \Delta
E_{1}^{m}} & , & m\neq 0\text{ and }m\neq N \\
1 & , & m=0,N%
\end{array}%
\right. \label{Bfm*-odd}
\end{equation}
for odd $N$, and
\begin{equation}
f_{m}^{\ast }=\left\{
\begin{array}{lll}
\frac{C_{m+2s}^{2s}C_{N-m+2s}^{2s}}{C_{N+2s}^{N}}e^{-\mu \beta \Delta
E_{1}^{m}} & , & m\neq 0,\frac{N}{2}\text{and }N \\
1 & , & m=0,N \\
\frac{\left( C_{N/2+2s}^{2s}\right) ^{2}}{C_{N+4s+1}^{N}} & , & m=\frac{N}{2}%
\end{array}%
\right. \label{Bfm*-even}
\end{equation}
\end{subequations}
for even $N$, where $\Delta E_{1}^{m}=\left\vert E_{1}\left( l_{eq}^{m}\right)
-E_{1}\left( L-l_{eq}^{m}\right) \right\vert $, $\mu =\min \left\{
m,N-m\right\} $ denotes the smaller of the number of particles on the first
level of both sides.

Thus we can obtain the total work in a thermodynamic cycle in  SQSZE

\begin{equation}
W_{tot}=\left\{
\begin{array}{lll}
k_{B}T\ln \left( \frac{C_{N+4s+1}^{4s+1}}{C_{N+2s}^{N}}\right)
-2\sum\limits_{m=1}^{\frac{N-1}{2}}m\frac{C_{m+2s}^{2s}C_{N-m+2s}^{2s}}{%
C_{N+4s+1}^{4s+1}}\Delta E_{1}^{m} & , & N\text{ is odd} \\
k_{B}T\left[ 1-\frac{\left( C_{N/2+2s}^{2s}\right) ^{2}}{C_{N+4s+1}^{N}}%
\right] \ln \left( \frac{C_{N+4s+1}^{4s+1}}{C_{N+2s}^{N}}\right)
-2\sum\limits_{m=1}^{\frac{N}{2}-1}m\frac{C_{m+2s}^{2s}C_{N-m+2s}^{2s}}{%
C_{N+4s+1}^{4s+1}}\Delta E_{1}^{m} & , & N\text{ is even}%
\end{array}%
\right. \label{Bwork}
\end{equation}

Similar to the fermion case, the total work in Eq. (\ref{Bwork}) can also be written as

\begin{equation}
W_{tot}=D_{B}k_{B}T-W_{0B} \label{Bworkexp}
\end{equation}
where
\begin{equation}
D_{B}=\frac{1}{k_{B}}\left( \frac{dW_{tot}}{dT}\right)
=\left\{
\begin{array}{lll}
\ln \left( \frac{C_{N+4s+1}^{4s+1}}{C_{N+2s}^{N}}\right)
 & \text{for} & N\text{ is odd} \\
\left[ 1-\frac{\left( C_{N/2+2s}^{2s}\right) ^{2}}{C_{N+4s+1}^{N}}%
\right] \ln \left( \frac{C_{N+4s+1}^{4s+1}}{C_{N+2s}^{N}}\right)
 & \text{for} & N\text{ is even}%
\end{array}%
\right.
\end{equation}
is proportional to the change rate of total work as temperature increases and $W_{0B}$ represents the absorbed work of boson SQSZE in the low temperature approximation.

\section{Basic properties of $D_{F},W_{0F},D_{B}$ and $W_{0B}$}
\subsection{Fermion SQSZE}
\begin{figure}
\includegraphics[width=4.8 in]{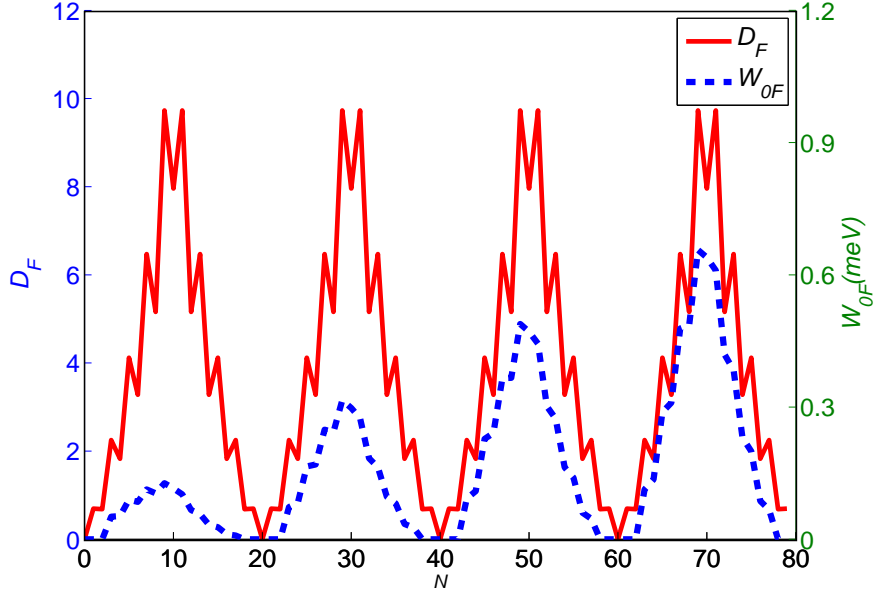}
\caption{\label{fig3}
(Color online) The change of $D_F$ and $W_{0F}$ along with $ N $, in which $ u $=5, $L=10^{-9} m$, $M=10^{-26} kg$
}
\end{figure}

For the given width of the quantum well $L=1nm$, the mass of particle $M=10^{-26}kg$, and the particle spin $s=9/2$,
we investigate $D_{F}$ and $W_{0F}$ versus $N$, which is shown in Fig.\ref{fig3}.
It can be seen that $D_{F}$, which represents the change rate of total work as temperature increases, shows even-odd oscillation
and periodicity, which modifies the parity effect of the spinless fermion QSZE\cite{lu}.
Besides, there also exists absorbed work in fermion SQSZE, which does not happen in spinless fermion QSZE. These behaviors of $W_{0F}$ and $D_{F}$
imply quantum nature of fermion SQSZE in low temperature.
In high temperature, both periodicity and even-odd oscillation of $W_{0F}$ and $D_{F}$ will vanish\cite{lu}.

Notice that the absorbed work increases as the particle number increases, we define the average absorbed work by

\begin{equation}
\overline{W_{0F}}=\frac{W_{0F}}{N} \label{Averagework}
\end{equation}

\begin{figure}
\includegraphics[width=5 in]{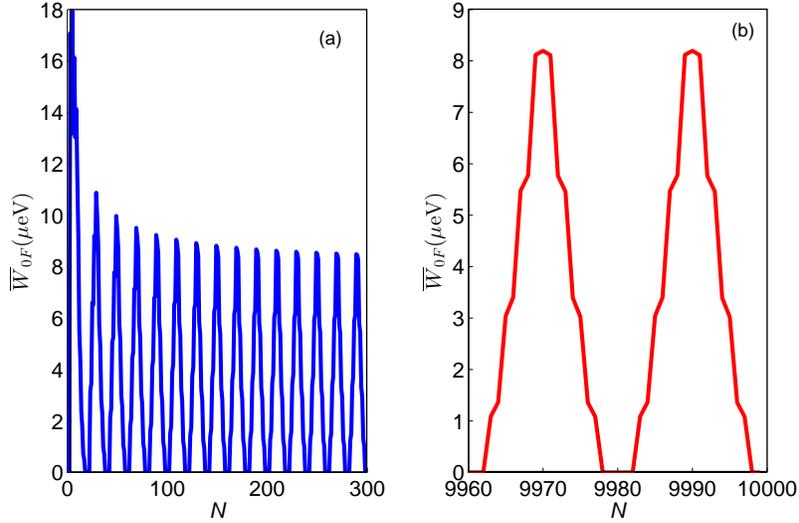}
\caption{\label{fig4}
(Color online) (a) The change of $\overline{W_{0F}}$ along with $N$. (b) The partial enlarged diagram of (a) when $N$ is very big. $ u $=5, $L=10^{-9} m$, $M=10^{-26} kg$}
\end{figure}

We investigate the average absorbed work versus the number of particles in the Fig.\ref{fig4}(a).
The $\overline{W_{0F}}$ declines rapidly with oscillation in the range of small number particles. As the number of particles increases,
the $\overline{W_{0F}}$ becomes periodic oscillation and trends to exact periodicity in the large-number limit of particles, as shown in Fig.\ref{fig4}(b). (See Appendix C)
This implies that the average absorbed work of fermion SQSZE is neither an intensive quantity nor an extensive quantity, which is also an unique characteristics of spin fermi system.

\begin{figure}
\includegraphics[width=5 in]{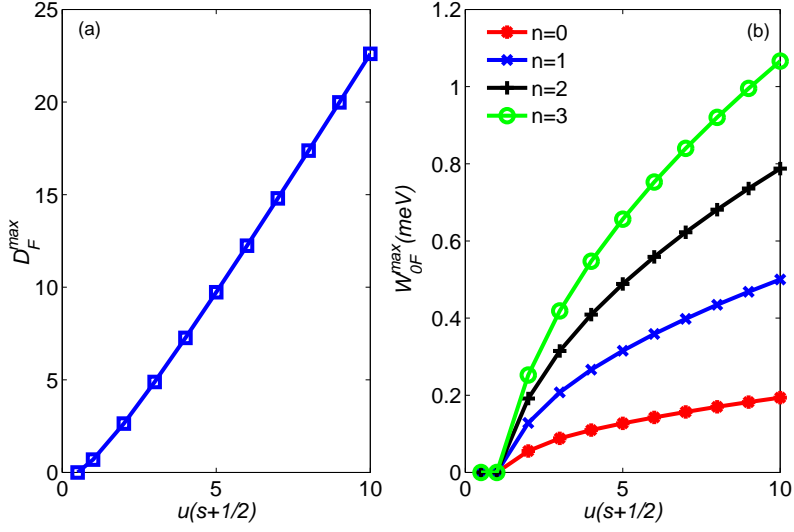}
\caption{\label{fig5}
(Color online) (a) The change of the maximum of $D_F$ along with $ u(u=s+1/2) $. (b) The change of the maximum of $W_{0F}$ along with $u$ when $n$ has different values. $ u $=5, $L=10^{-9} m$, $M=10^{-26} kg$}
\end{figure}

We can also investigate the maximum of $D_{F}$ and $W_{0F}$ versus spin $u=s+1/2$ of particles shown in Fig. \ref{fig5}.
The maximum of $D_{F}$ and $W_{0F}$ increases
as spin of particles grows. It implies that the spin of particles plays an important role in the total work of fermion SQSZE, which is also a quantum nature of of fermion SQSZE.\\

\begin{figure}
\includegraphics[width=5 in]{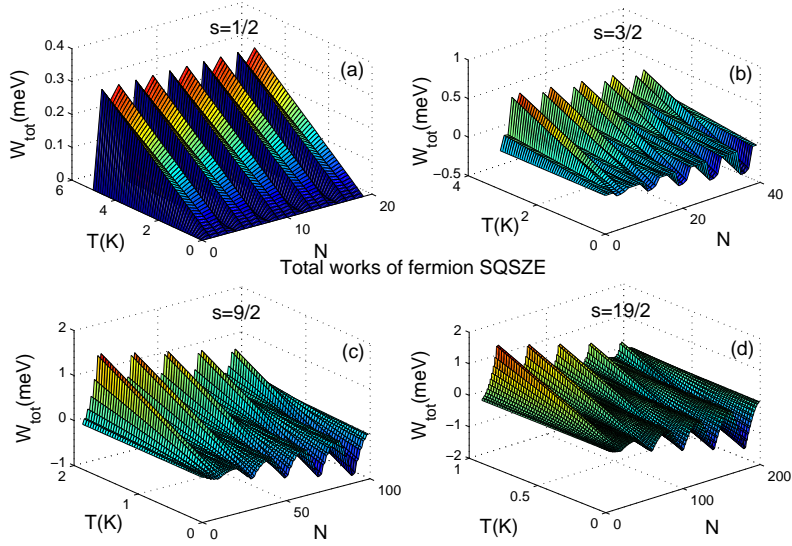}
\caption{\label{fermiwork}
(Color online) The change of total work $W_{tot}$ along with temperature $T$ and particle number $N$ when (a) $u$=1, (b)$u$=2, (c)$u$=5, (d)$u$=10. $L=10^{-9} m$, $M=10^{-26} kg$
}
\end{figure}

As shown in Fig.\ref{fermiwork}, we can explicitly plot the relationships between the total work $W_{tot}$, temperature $T$ and the particle number $N$. It can be seen that the total works of the different-spin fermion SQSZE oscillate with the particle number increasing, and increase with temperature. For $s=1/2$ case, the total work $W_{tot}\ge 0$ in the whole $T-N$ region, which implies that there is no absorbed work. For $s>1/2$ cases, the total works can be negative in the large-N regions, which can be viewed as a phase transition induced by the absorbed work variation in the high-spin fermion SQSZE, which can be also seen in the phase diagram in Fig. \ref{fig7}.

\subsection{Boson SQSZE}
\begin{figure}
\includegraphics[width=5 in]{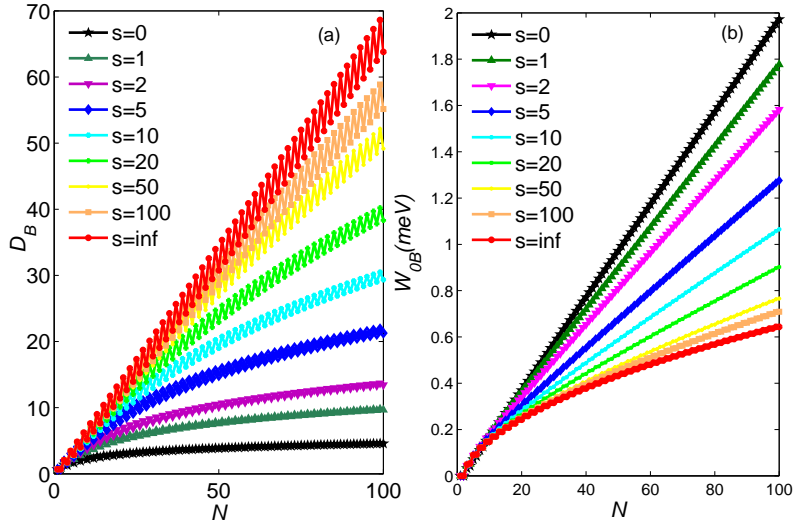}
\caption{\label{fig6}
(Color online) The change of $D_B$ and $W_{0B}$ along with $ N $ when $s$ gets different values, in which $L=10^{-9} m$, $M=10^{-26} kg$}
\end{figure}

For boson SQSZE, the $D_{B}$ and $W_{0B}$ versus $N$ for given different spin $s$ is shown in Fig. \ref{fig6}.
It can be seen that both $D_{B}$ and $W_{0B}$ increase with the particle number, and
there is some oscillations when the spin increases. Interestingly, $D_{B}$ increase with spin increasing, but $W_{0B}$ decreases with spin increasing.
These properties reveal the difference between boson and fermion SQSZE.

In the large spin limit, $D_{B}$ and $W_{0B}$ can be expressed as
\begin{equation}
\lim_{s\rightarrow \infty }D_{B}=\left\{
\begin{array}{lll}
N\ln 2 & \text{for} & N\text{ is odd} \\
\left( 1-\frac{C_{N}^{N/2}}{2^{N}}\right) N\ln 2 & \text{for} & N\text{ is even}%
\end{array}%
\right. \label{BDlimit}
\end{equation}

\begin{equation}
\lim_{s\rightarrow \infty }W_{0B}=\left\{
\begin{array}{lll}
\frac{1}{2^{N-1}}\sum_{m=1}^{\frac{N-1}{2}}mC_{N}^{m}\Delta E_{1}^{m} & \text{for} & N%
\text{ is odd} \\
\frac{1}{2^{N-1}}\sum_{m=1}^{\frac{N}{2}-1}mC_{N}^{m}\Delta E_{1}^{m} & \text{for} & N%
\text{ is even}%
\end{array}%
\right. \label{BWlimit}
\end{equation}

\begin{figure}
\includegraphics[width=5 in]{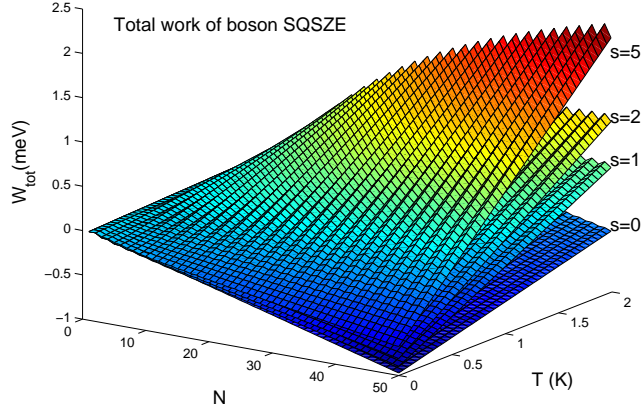}
\caption{\label{BosonWork}
(Color online) The change of total work $W_{tot}$ along with temperature $T$ and particle number $N$ when $s$=0,1,2,5. $L=10^{-9} m$, $M=10^{-26} kg$
}
\end{figure}

To explicitly understand the relationships between the total work $W_{tot}$, temperature $T$ and particle number $N$, we plot the Fig.\ref{BosonWork}. It can be found that the total work $W_{tot}$ of the boson SQSZE increases with temperature and spin increasing, but has no great oscillation with the particle number. There is a low-temperature region, in which the total work is negative. As spin increases, this region becomes small, which can be seen clearly from the phase diagram in Fig. \ref{fig7}. The basic behavior of the total work of the boson SQSZE is quite different from fermi SQSZE, especially for high-spin working matters.

\subsection{Phase diagram of SQSZE}
As a heat engine, an important issue is the relationship between work, temperature and particle number. We define the critical temperature of fermion SQSZE to study the working properties of SQSZE,

\begin{equation}
T_{cF}=\frac{W_{0F}}{D_{F}k_{B}}, where \  N\neq 4un \label{FTcdef}
\end{equation}

\begin{figure}
\includegraphics[width=4.8 in]{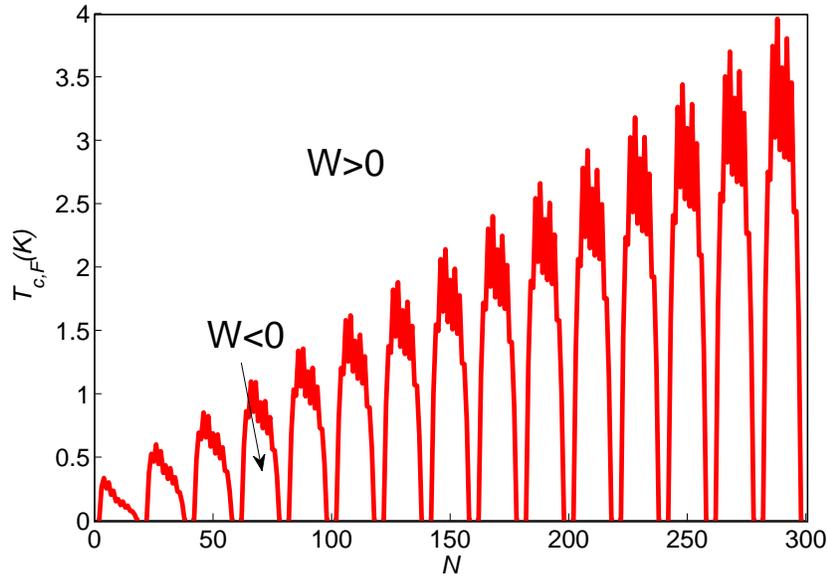}
\caption{\label{fig7}
(Color online) The work phase diagram of the fermion QSZE, in which $ u $=5, $L=10^{-9} m$, $M=10^{-26} kg$.}
\end{figure}

We numerically investigate the working properties of fermion SQSZE in terms of the critical temperature and the particle number in Fig. \ref{fig7},
which can be regarded as a phase diagram of fermion SQSZE.
The curve in the phase diagram goes up and shows a periodic oscillation.
The upper region of the curve in the phase diagram denotes the fermion SQSZE engine doing positive work, while the lower region of the curve represents the fermion SQSZE doing negative work.

Similarly, the critical temperature of the boson SQSZE can also be defined by

\begin{equation}
T_{cB}=\frac{W_{0B}}{D_{B}k_{B}}, where \ N\neq 0 \label{BTcdef}
\end{equation}

\begin{figure}
\includegraphics[width=4.8 in]{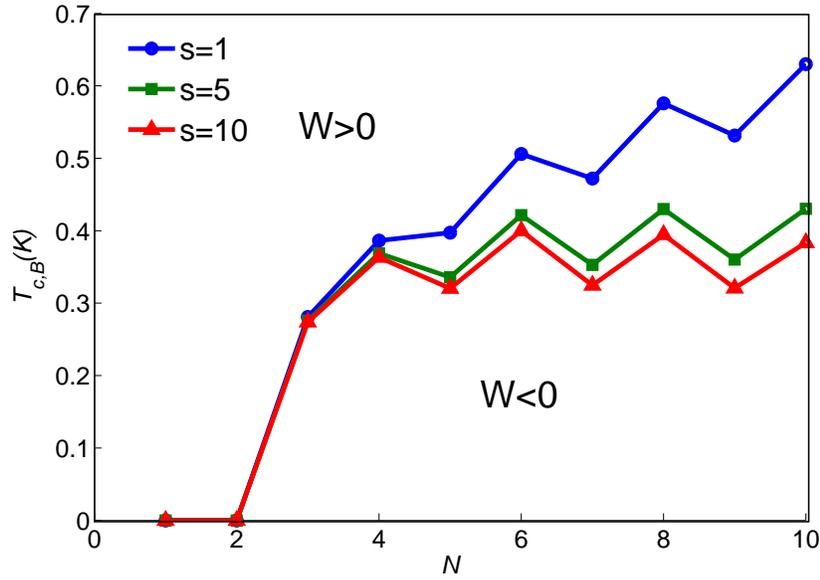}
\caption{\label{fig8}
(Color online) The work phase diagram of the boson QSZE when $s$ gets different values, in which $L=10^{-9} m$, $M=10^{-26} kg$.}
\end{figure}

The phase diagram of the boson SQSZE is plotted in Fig. \ref{fig8} for given spin of particles.
The critical curves for $W=0$ go up with oscillation and separate for different spins as the number of particles increases, but it looks quite different from the curve of the fermion SQSZE.

Actually, Lu and Long found that the phase transition occurs in the spinless boson SQSZE\cite{lu}, and there have been some works discussing the superposition effect in a two-state quantum engine\cite{ou1,ou2}, its effeciency\cite{Yan,Wang}. However, all of these works ignore the spin effect in the quantum engine. Here, we found that the spin degree of freedom can decrease the critical temperature of the phase transition in boson SQSZE. This is because the spin degree of freedom will modify the configuration states of the working matter.
\section{Information-erasure work and information-work efficiency}

According to the Landauer's erasure principle, the erasure or resetting of
memory results in the energy dissipation, which would cause entropy increase.
The erasure work is proportional to the amount of information stored in memory,\cite{landauer}

\begin{equation}
W_{eras}=k_{B}T\ln 2H(P) \label{Werasure}
\end{equation}
where $H(P)=-\sum_{i=1}^{n}P_{i}\log _{2}P_{i}$ represents Shannon entropy
(or information entropy) in unit Bit.
Due to the quantum indistinguishability of particles, the demon does not know where particles are.
In fact, the information that demon needs to acquire in the measurement process is the number of identical particles on each side. Since $f_{m}$ is the probability of finding $m$ particles on
the left side, the erasure work is given by

\begin{equation}
W_{eras}=-k_{B}T\ln 2\sum_{m=0}^{N}f_{m}\log
_{2}f_{m}=-k_{B}T\sum_{m=0}^{N}f_{m}\ln f_{m} \label{WeraQSZE}
\end{equation}

Suppose that the erasure and resetting of the memory happen in every thermodynamic cycle, and the erasure
work is provided by the engine. The net work in a thermodynamic cycle of SQSZE can do is given by

\begin{equation}
W_{net}=W_{tot}-W_{eras}=k_{B}T\sum_{m=0}^{N}f_{m}\ln f_{m}^{\ast } \label{Wnet}
\end{equation}

It can be seen that the profound relationship between the net work in Eq.(\ref{Wnet}) and the second law of thermodynamics. Since $f_{m}^{\ast }\leq 1$, the net work done by
SQSZE must be negative or zero, which corresponds with the Kelvin's statement
of the second law of thermodynamics because the heat engine cannot extract any
work from a single heat source in a thermodynamic cycle, after resetting the whole system including
the memory to the initial state. This irreversibility can also be understood by the tunnelling effect,
which leads to the irreversibility as diffusion does. When $f_{m}^{\ast }=1$ for given $m$ , i.e $Z_{m}(l_{eq}^{m})=Z(l_{eq}^{m})$, there is no
tunnelling or tunneling does not take effect for the case that $m$ particles are on the left side, so unless all the $f_{m}^{\ast }$ equal to $1$, the SQSZE must be irreversible due to the tunnelling effect.

We hope that the work done by SQSZE is as much as possible while
the erasure work is as little as possible. In order to judge the quality of
a SQSZE from this aspect, we define an efficiency as the ratio of the total work and the erasure work, we called it as the information-work efficiency.

\begin{equation}
\eta_{i-w}=\frac{W_{tot}}{W_{eras}}
\label{Effdef}
\end{equation}

Notice that the total work done by QSZE might be negative, the domain of information-work efficiency is $\left( -\infty ,1\right] $. When the information-work efficiency is negative, it means that the engine cannot do any work. When the information-work
efficiency is positive, it implies the engine can do work
even though it still needs erasure work.
When the information-work efficiency
is higher, we believe that the engine is more efficient and more reversible
because less energy is dissipated. In particular, the engine
become reversible when the information-work efficiency reaches $1$ because
there is no energy dissipation.

Substitute Eqs.(\ref{Ffm-1})(\ref{Ffm-2}) (\ref{Fworkexp}) and(\ref{WeraQSZE}) into Eq.(\ref{Effdef}), we obtain the information-work efficiency in fermion system

\begin{equation}
\eta _{i-w}^F=\left\{
\begin{array}{lll}
\frac{D_{F}k_{B}T-W_{0F}}{-k_{B}T\sum_{p=0}^{k}f_{p}\ln f_{p}}
 & \text{for} & 0\leq k<2u \\
\frac{D_{F}k_{B}T-W_{0F}}{-k_{B}T\sum_{q=0}^{4u-k}f_{q}\ln f_{q}}
 & \text{for} & 2u\leq k<4u%
\end{array}%
\right. \label{Feff}
\end{equation}
where $f_{p} =\frac{C_{2u}^{p}C_{2u}^{k-p}}{C_{4u}^{k}}$ and
$f_{q}=\frac{C_{2u}^{q}C_{2u}^{4u-k-q}}{C_{4u}^{4u-k}}$.
Likewise, the information-work efficiency in boson system is given by

\begin{equation}
\eta _{i-w}^B=\frac{D_{B}k_{B}T-W_{0B}}{-k_{B}T\sum_{m=0}^{N}
f_{m}\ln f_{m}}  \label{Beff}
\end{equation}
where $f_{m} =\frac{C_{m+2s}^{2s}C_{N-m+2s}^{2s}}{C_{N+4s+1}^{4s+1}}$.

\section{The condition of maximum extracted work and highest information-work efficiency}

Naturally, we want to know the condition of maximum extracted work and highest information-work efficiency. From Eqs.(\ref{Fworkexp}) and (\ref{Bworkexp}) we know that the extracted work can reach maximum when the engine does not absorb work from environment, namely $W_{0F}=0$ (or $W_{0B}=0$). It can be verified that $W_{0F}=0$ only when $k=0,1,2,4u-2,4u-1$ and $W_{0B}=0$ when $N=0,1,2$. 
So we can verify that, for $k=1,4u-1$ the maximum extracted work of fermion SQSZE is $W_{tot}^{F,max }=k_{B}T\ln 2$, and for $N=2$ the maximum extracted work of boson SQSZE is $W_{tot}^{B,max }=k_{B}T\frac{2s+2}{4s+3}\ln \frac{4s+3}{s+1}$.
Interestingly, $W_{tot}^{B,max }=k_{B}T\frac{2s+2}{4s+3}\ln \frac{4s+3}{s+1}>k_{B}T\ln 2$, which can be generalized to $W_{tot}^{B,max }>W_{tot}^{F,max }$ for all cases. This means that we should use bosons as working substances in order to obtain maximum extracted work.

For the information-work efficiency defined by Eq.(\ref{Effdef}), it is possible to reach
maximum when the engine does positive work, i.e. $k = 0,1,2,4u-2,4u-1$ or $N = 0,1,2$. By using the Eqs. (\ref{Feff}), for fermion SQSZE, the maximum information-work
efficiency can be obtained $\eta_{i-w}^{F,max}=1$ for $k=1,4u-1$.
Similarly, for boson SQSZE, the information-work efficiency reaches maximum,
$\eta _{i-w}^{B,max}=1$ for $N=1$.

Interestingly, the second-highest information-work efficiency of both fermion and boson SQSZEs
can be expressed as a same form,
\begin{equation}
\eta_{i-w}^{F(B),h2}=\frac{1}{1+\frac{(1-\alpha _{F(B)})\ln (1-\alpha _{F(B)})}{%
\alpha _{F(B)}\ln \frac{\alpha _{F(B)}}{2}}} \label{Feff2max}
\end{equation}
where $\alpha _{F}=\frac{2u-1}{4u-1}=\frac{2s}{4s+1}$ with $k=2$ or $4u-2$ for fermion SQSZE,
and $\alpha _{B}=\frac{2s+2}{4s+3}$ with $N=2$ for boson SQSZE.

Therefore, the working particle number plays an important role in the extracted work and information-work efficiency for SQSZE. If we can control the particle number in $4un+1$ or $4un+4u-1$, where $n\in \mathbb{N}$, the fermion SQSZE has the maximum extracted work and highest information-work efficiency.
However, for the boson SQSZE, when the working particle number is $N=2$, the boson SQSZE has maximum extracted work and the highest
information-work efficiency corresponds to the working particle number $N=1$.
Moreover, when $\eta _{i-w}=1$, the thermodynamic cycle is reversible, namely, when $N=4un+1$, or $4un+4u-1$ the fermi SQSZE  and $N=1$  for the boson SQSZE, the QSZE becomes reversible.

\section{conclusion}
The QSZE plays an important role in understanding the fundamental physics of the quantum engine, information, and quantum thermodynamics. We generalize QSZE to spin QSZE (SQSZE) and systematically study the basic physical properties of both fermion and boson SQSZEs in low temperature approximation. We give the analytic formulation of the total work, information erasure work, and efficiency. Based on this formulation, we find a series of the physical properties of SQSZE: \\
(1) The total work depends on both particle number and particle spin. The change rate $D_{F}$ as temperature increases exhibits periodicity and even-odd oscillation with the particle number. The parity effect embeds in the periodicity and even-odd oscillation. This is a generalization of spinless fermion QSZE, in which the total work depends on only particle parity.\cite{lu}\\
(2) For the fermion SQSZE, the average absorbed work from environment osicillates regularly and periodically in large-number limit, which implies that the average absorbed work
is neither an intensive quantity nor an extensive quantity. This is a novel and pure quantum phenomenon in quantum thermodynamics.
\\
(3) The phase diagrams of both fermion and boson SQSZEs gives the SQSZE doing positive or negative work in the parameter space of the critical temperature and the particle number of system. The critical temperature of phase transition depends sensitively on the particle number.\\
(4) By using the Landauer's erasure principle, we give the erasure work in a thermodynamic cycle and define a new efficiency (we refer it information-work efficiency) to measure the engine's ability of utilizing information to extract work. \\
(5) We give the particle number conditions under which the maximum extracted work and highest information-work efficiencies for fermion and boson SQSZEs can be achieved.

These results reveal the basic physics of SQSZE, especially for characteristic difference between spinless QSZE and spin QSZE, as well as fermion and boson working particles in SQSZE.
These physical properties of SQSZE provide some hints to understand the relationships between quantum mechanics, quantum thermodynamics and information physics.

\appendix
\section{DERIVATIONS OF $f_{m}$ and $f_{m}^{\ast }$ IN FERMI SYSTEM}
Let us consider $N$ fermions with spin quantum number $s$ occupying the energy levels. Defining $u=s+\frac{1}{2}$ and suppose that $N=4un+k$ where $n\geq 0$, we know that each  spin-$s$ fermion has $2u$ different spin states, which is equivalent to $2u$ degrees of
degeneracy on each energy level.

{\bf Derivation of $f_{m}$}

In low-temperature approximation, we ignore the temperature fluctuation and consider only the ground state in the canonical ensemble, namely, particles occupy as low energy levels as possible and are constrained by Pauli's exclusive principle.
Suppose that the engine system is an infinite potential well and in the first step of SQSZE, a wall is inserted at a half width position.
Thus, there are $4un$ particles occupying the low energy levels $1,2,...n$ in the
left and right sides of well, and the other $k$ particles occupy the $(n+1)$th energy level
in the left or right side. Namely, $k$ particles occupy two energy levels with $2s+1$ degrees of degeneracy .
Using the combinatorics, the number of configuration states is $C_{4u}^{k}$.
Therefore, the partition function $Z(\frac{L}{2})$ in the low-temperature approximation is given by

\begin{equation}
Z\left( \frac{L}{2} \right)= C_{4u}^{k}e^{-\beta\left[4u \sum_{i=1}^{n}E_{i}\left( \frac{L}{2}\right) +kE_{n+1}\left(\frac{L}{2}\right)\right]}
\label{FZ(L/2)}
\end{equation}
where $E_{\alpha}\left(\frac{L}{2}\right)=\frac{\alpha^{2}\pi^{2}\hbar^{2}}{2M(L/2)^{2}}$.
$Z_{m}(\frac{L}{2})$ is the partition function for $m$ particles on the left side.
For $0\leq k<2u$, the domain of $m$ is $2un\leq m\leq 2un+k$, and for $2u\leq k<4u$,
the domain of $m$ is $2un+k-2u\leq m\leq 2un+2u$, since if $m$ equals to other values, $Z_{m}(L/2)\ll Z(L/2)$ and $f_{m}\rightarrow0$, which means it is impossible for $m$ particles located on the left side. Suppose that $p=m-2un$ particles occupy the $(n+1)$th energy level on the left side, namely $k-p$ particles occupy the $(n+1)$th energy level on the right side of the well, the partition function $Z_{m}(\frac{L}{2})$ is given by

\begin{equation}
Z_{m}\left(\frac{L}{2}\right)= C_{2u}^{p}C_{2u}^{k-p}e^{-4u\beta
\sum_{i=1}^{n}E_{i}\left( \frac{L}{2}\right) -k\beta E_{n+1}\left( \frac{L}{2%
}\right) }
\label{FZm(L/2)}
\end{equation}
Notice that for $2u\leq k<4u$, it is convenient to express the number of configuration states in terms of the hole-representation, we obtain the probability

\begin{equation}
f_{m}=\frac{Z_{m}\left(\frac{L}{2}\right)}{Z\left( \frac{L}{2} \right)}=\left\{
\begin{array}{lll}
\frac{C_{2u}^{p}C_{2u}^{k-p}}{C_{4u}^{k}} & \text{for} & 0\leq k<2u   \text{ and } 0\leq p\leq k\\
\frac{C_{2u}^{q}C_{2u}^{4u-k-q}}{C_{4u}^{4u-k}} & \text{for} & 2u\leq k<4u  \text{ and } 0\leq q\leq 4u-k\\
0 & \text{for} & \text{otherwise}
\end{array}%
\right. \label{AFfm1}
\end{equation}
where $q=2u-p=2u(n+1)-m$ represents the number of unoccupied states on the $%
(n+1)$th energy level of the left side.

{\bf Derivation of $f_{m}^{\ast }$}

For $f_{m}^{\ast }$, the wall expands to a balance position $\ell^{m}_{eq}$. It satisfies the balance equation \cite{lu,callen}

\begin{equation}
\sum_{n=1}^{\infty }P_{n}\left( \ell_{eq}^{m}\right) \frac{\partial E_{n}\left(
\ell_{eq}^{m}\right) }{\partial \ell_{eq}^{m}}=\sum_{n=1}^{\infty }P_{n}\left(
L-\ell_{eq}^{m}\right) \frac{\partial E_{n}\left( L-\ell_{eq}^{m}\right) }{%
\partial \left( L-\ell_{eq}^{m}\right) } \label{walleqdef}
\end{equation}
where $P_{n}\left(\ell_{eq}^{m}\right)$ represents the average number of particles on the $n$th energy level on the left side, and $P_{n}\left(L-\ell_{eq}^{m}\right)$ on the right side. In the low-temperature approximation, particles occupy as low energy levels as possible. Thus, the Eq. (\ref{walleqdef}) can be reduced to

\begin{equation}
\left( \frac{\ell_{eq}^{m}}{L-\ell_{eq}^{m}}\right) ^{3}=\frac{2u\sum_{\xi =1}^{%
\left[ \frac{m}{2u}\right] }\xi ^{2}+\left( \left[ \frac{m}{2u}\right]
+1\right) ^{2}a }{2u\sum_{\xi =1}^{\left[ \frac{N-m}{2u}\right] }\xi
^{2}+\left( \left[ \frac{N-m}{2u}\right] +1\right) ^{2}b } \label{Fwalleq1}
\end{equation}
where $\left[ x\right]=max(i\in \mathbb{Z},i\leq x)$, and $a =\mod(m,2u)$ and $b =\mod(N-m,2u)$.
The Eq. (\ref{Fwalleq1}) can be simplified to

\begin{equation}
\left( \frac{\ell_{eq}^{m}}{L-\ell_{eq}^{m}}\right) ^{3}=\frac{un(2n+1)+3p(n+1)}{%
un(2n+1)+3(k-p)(n+1)} \label{Fwalleq2}
\end{equation}

It can be verified that $E_{n+1}(\ell_{eq}^{m})>E_{n}(L-\ell_{eq}^{m})$, and $E_{n}(\ell_{eq}^{m})<E_{n+1}(L-\ell_{eq}^{m})$, so the $2n$ lowest degenerate energy levels of the systems are $E_{1}(\ell_{eq}^{m})E_{2}(\ell_{eq}^{m})\cdots E_{n}(\ell_{eq}^{m})$ and
$E_{1}(L-\ell_{eq}^{m})E_{2}(L-\ell_{eq}^{m})\cdots E_{n}(L-\ell_{eq}^{m})$.

{\bf Case (1)}: $0\leq k<2u$

When $k$ is odd and $p<\frac{k}{2}$, the partition function of the system for the wall at the balance $\ell_{eq}^{m}$ is

\begin{equation}
Z\left( \ell_{eq}^{m}\right) = C_{2u}^{k}e^{-\beta \left[ 2u\left(
\sum_{i=1}^{n}E_{i}(\ell_{eq}^{m})+\sum_{i=1}^{n}E_{i}(L-\ell_{eq}^{m})\right)
+kE_{n+1}(L-\ell_{eq}^{m})\right] } \label{FZ(leq)g}
\end{equation}
and similarly, the partition function for $m$ particles on the left side is obtained
\begin{equation}
Z_{m}\left( \ell_{eq}^{m}\right) = C_{2u}^{p}C_{2u}^{k-p}e^{-\beta \left[
2u\left(
\sum_{i=1}^{n}E_{i}(\ell_{eq}^{m})+\sum_{i=1}^{n}E_{i}(L-\ell_{eq}^{m})\right)
+pE_{n+1}(\ell_{eq}^{m})+\left( k-p\right) E_{n+1}(L-\ell_{eq}^{m})\right] } \label{FZm(leq)g}
\end{equation}
Consequently, the probability is obtained
\begin{equation}
f_{m}^{\ast }=\frac{Z_{m}\left( \ell_{eq}^{m}\right)}{Z\left( \ell_{eq}^{m}\right)}=\frac{C_{2u}^{p}C_{2u}^{k-p}}{C_{2u}^{k}}e^{-p\beta \left(
E_{n+1}(\ell_{eq}^{m})-E_{n+1}(L-\ell_{eq}^{m})\right) } \label{AFfm*1-odd-1}
\end{equation}
When $k$ is even and $p>\frac{k}{2}$, similarly to the Eq. (\ref{AFfm*1-odd-1}), we
obtain

\begin{equation}
f_{m}^{\ast }=\frac{C_{2u}^{p}C_{2u}^{k-p}}{C_{2u}^{k}}e^{-\left( k-p\right)
\beta \left( E_{n+1}(L-\ell_{eq}^{m})-E_{n+1}(\ell_{eq}^{m})\right) } \label{AFfm*1-odd-2}
\end{equation}

Combining the Eq. (\ref{AFfm*1-odd-1}) and Eq. (\ref{AFfm*1-odd-2}), the $f_{m}^{\ast}$ can be rewritten as

\begin{equation}
f_{m}^{\ast }=\frac{C_{2u}^{p}C_{2u}^{k-p}}{C_{2u}^{k}}e^{-\mu \beta \Delta
E_{n+1}^{m}} \label{AFfm*1-odd}
\end{equation}
where $\Delta E_{n+1}^{m}=\left\vert E_{n+1}\left( \ell_{eq}^{m}\right)
-E_{n+1}\left( L-\ell_{eq}^{m}\right) \right\vert $, $\mu =\min \left\{
p,k-p\right\} $ denotes the smaller one of the number of particles on the $%
(n+1)$th level of the both sides.

However, for $0\leq k<2u$, and $k$ is even, $p=\frac{k}{2}$,
$\ell_{eq}^{m}=L-\ell_{eq}^{m}=\frac{L}{2}$, the $(n+1)$th energy level of both sides
is the same, we get

\begin{equation}
Z\left( l_{eq}^{m}\right) = C_{4u}^{k}e^{-\beta \left[ 4u\sum_{i=1}^{n}E_{i}%
\left( \frac{L}{2}\right) +kE_{n+1}\left( \frac{L}{2}\right) \right] } \label{FZ(leq)s}
\end{equation}
and
\begin{equation}
Z_{m}\left( l_{eq}^{m}\right) = C_{2u}^{\frac{k}{2}}C_{2u}^{\frac{k}{2}%
}e^{-\beta \left[ 4u\sum_{i=1}^{n}E_{i}\left( \frac{L}{2}\right)
+kE_{n+1}\left( \frac{L}{2}\right) \right] } \label{FZm(leq)s}
\end{equation}
Consequently, we obtain
\begin{equation}
f_{m}^{\ast }=\frac{C_{2u}^{\frac{k}{2}}C_{2u}^{\frac{k}{2}}}{C_{4u}^{k}} \label{AFfm*1-even-2}
\end{equation}

{\bf Case (2)}: $2u\leq k<4u$

Similar to the case (1), when $k$ is odd, we get

\begin{equation}
f_{m}^{\ast }=\frac{C_{2u}^{q}C_{2u}^{4u-q-k}}{C_{2u}^{4u-k}}e^{-\lambda
\beta \Delta E_{n+1}^{m}} \label{AFfm*2-odd}
\end{equation}
and when $k$ is even and $q\neq \frac{4u-k}{2}$, we have
\begin{subequations}
\label{AFfm*2-evenwhole}
\begin{equation}
f_{m}^{\ast }=\frac{C_{2u}^{q}C_{2u}^{4u-q-k}}{C_{2u}^{4u-k}}e^{-\lambda
\beta \Delta E_{n+1}^{m}} \label{AFfm*2-even1}
\end{equation}

When $k$ is even, but $q=\frac{4u-k}{2}$
\begin{equation}
f_{m}^{\ast }=\frac{C_{2u}^{\frac{4u-k}{2}}C_{2u}^{\frac{4u-k}{2}}}{%
C_{4u}^{4u-k}} \label{AFfm*2-even2}
\end{equation}
\end{subequations}
where $\lambda =\min \left\{ 4u-q-k,q\right\}$ denotes the smaller of the
number of unoccupied states on the $(n+1)$th level of the both sides.

In particular, when $n=0$ and $p=0,k$, the wall is moved to the end of box
so $f_{m}^{\ast }=1$.

\section{DERIVATIONS OF $f_{m}$ and $f_{m}^{\ast }$ IN BOSE SYSTEM}
Suppose that the system has $N$ bosons with spin quantum number $s$, we know that each boson has $2s+1$ different spin states, which leads to $2s+1$ degrees of degeneracy on each energy level.
In the low temperature approximation, we also ignore the temperature fluctuation and consider only the ground state in the canonical ensemble.

{\bf Derivation of $f_{m}$}

Notice that when the number of particles is $a$ and the degrees of
degeneracy is $w$, the number of micro-states is given by $\Omega=C_{w+a-1}^{a}$ \cite{callen}, we get

\begin{equation}
Z\left( \frac{L}{2}\right) = C_{N+4s+1}^{4s+1}e^{-N\beta E_{1}\left( \frac{L}{%
2}\right) }
\label{BZ(L/2)}
\end{equation}
and
\begin{equation}
Z_{m}\left( \frac{L}{2}\right)= C_{m+2s}^{2s}C_{N-m+2s}^{2s}e^{-N\beta
E_{1}\left( \frac{L}{2}\right) }
\label{BZm(L/2)}
\end{equation}
Therefore, we obtain
\begin{equation}
f_{m}=\frac{Z_{m}\left( \frac{L}{2}\right)}{Z\left( \frac{L}{2}\right)}=\frac{C_{m+2s}^{2s}C_{N-m+2s}^{2s}}{C_{N+4s+1}^{4s+1}} \label{ABfm}
\end{equation}

{\bf Derivation of $f_{m}^{\ast }$}

In the boson case, the Eq. (\ref{walleqdef}) becomes

\begin{equation}
\left( \frac{\ell_{eq}^{m}}{L-\ell_{eq}^{m}}\right) ^{3}=\frac{m}{N-m} \label{Bwalleq}
\end{equation}

Since the lowest energy level of boson system is $E_{1}(\ell_{eq}^{m})$ or $%
E_{1}(L-\ell_{eq}^{m})$.
When $N$ is odd and $0<m\leq \frac{N-1}{2}$, we get

\begin{equation}
Z\left( \ell_{eq}^{m}\right)=C_{N+2s}^{N}e^{-N\beta E_{1}\left(
L-\ell_{eq}^{m}\right) }
\label{BZ(leq)-1}
\end{equation}
and
\begin{equation}
Z_{m}\left( \ell_{eq}^{m}\right)=C_{m+2s}^{2s}C_{N-m+2s}^{2s}e^{-\beta \left[
mE_{1}\left( \ell_{eq}^{m}\right) +\left( N-m\right) E_{1}\left(
L-\ell_{eq}^{m}\right) \right] }
\label{BZm(leq)-1}
\end{equation}
Consequently, we obtain
\begin{equation}
f_{m}^{\ast }=\frac{Z_{m}\left( \ell_{eq}^{m}\right)}{Z\left( \ell_{eq}^{m}\right)}=\frac{C_{m+2s}^{2s}C_{N-m+2s}^{2s}}{C_{N+2s}^{N}}e^{-m\beta %
\left[ E_{1}\left( \ell_{eq}^{m}\right) -E_{1}\left( L-\ell_{eq}^{m}\right) \right]
} \label{ABfm*1}
\end{equation}

Similarly,when $N$ is odd and $\frac{N+1}{2}\leq m<N$, we can obtain

\begin{equation}
f_{m}^{\ast }=\frac{C_{m+2s}^{2s}C_{N-m+2s}^{2s}}{C_{N+2s}^{N}}e^{-\left(
N-m\right) \beta \left[ E_{1}\left( L-\ell_{eq}^{m}\right) -E_{1}\left(
\ell_{eq}^{m}\right) \right] } \label{ABfm*2}
\end{equation}

Combining the Eqs. (\ref{ABfm*1}) and (\ref{ABfm*2}), we get

\begin{equation}
f_{m}^{\ast }=\frac{C_{m+2s}^{2s}C_{N-m+2s}^{2s}}{C_{N+2s}^{N}}e^{-\mu \beta
\Delta E_{1}^{m}} \label{ABfm*odd}
\end{equation}
where $\Delta E_{1}^{m}=\left\vert E_{1}\left( \ell_{eq}^{m}\right)
-E_{1}\left( L-\ell_{eq}^{m}\right) \right\vert $, $\mu =\min \left\{
m,N-m\right\} $ denotes the smaller of the number of particles on the first
level of the both sides.

By the same way, when $N$ is even and $m\neq \frac{N}{2},0,N$, we get

\begin{equation}
f_{m}^{\ast }=\frac{C_{m+2s}^{2s}C_{N-m+2s}^{2s}}{C_{N+2s}^{N}}e^{-\mu \beta
\Delta E_{1}^{m}} \label{ABfm*even1}
\end{equation}

However, when $N$ is even and $m=\frac{N}{2}$, $\ell_{eq}^{m}=L-\ell_{eq}^{m}=%
\frac{L}{2}$, the first energy level of both sides is the same, so we get

\begin{equation}
f_{m}^{\ast }=\frac{C_{\frac{N}{2}+2s}^{2s}C_{\frac{N}{2}+2s}^{2s}}{%
C_{N+4s+1}^{N}} \label{ABfm*even2}
\end{equation}

In particular, when $m=0$ or $m=N$, the wall reaches the end of the box,
which makes $f_{m}^{\ast }=1$.

\section{DERIVATION OF $\lim_{n\rightarrow \infty }\overline{W_{0F}}$}
For the infinite potential well model, $\Delta E_{n+1}^{m}$ is written as

\begin{equation}
\Delta E_{n+1}^{m}=\frac{\left( n+1\right) ^{2}\pi ^{2}\hbar ^{2}}{2M}%
\left\vert \frac{1}{\left( \ell_{eq}^{m}\right) ^{2}}-\frac{1}{\left(
L-\ell_{eq}^{m}\right) ^{2}}\right\vert \label{C1}
\end{equation}

It can be rewritten as
\begin{equation}
\Delta E_{n+1}^{m}=\frac{\left( n+1\right) ^{2}\pi ^{2}\hbar ^{2}}{2ML^{2}}%
\cdot \frac{\left( 1+\frac{l_{eq}^{m}}{L-l_{eq}^{m}}\right) ^{3}}{\left(
\frac{l_{eq}^{m}}{L-l_{eq}^{m}}\right) ^{2}\left[ 1+\frac{l_{eq}^{m}}{%
L-l_{eq}^{m}}+\left( \frac{l_{eq}^{m}}{L-l_{eq}^{m}}\right) ^{2}\right] }%
\cdot \left\vert 1-\left( \frac{l_{eq}^{m}}{L-l_{eq}^{m}}\right)
^{3}\right\vert \label{C2}
\end{equation}
By substituting Eq. (\ref{Fwalleq2}) to Eq. (\ref{C2}), we get

\begin{equation}
\Delta E_{n+1}^{m}\approx \frac{4\pi ^{2}\hbar ^{2}}{3ML^{2}}\left(
n+1\right) ^{2}\left\vert \frac{3\left( k-2p\right) \left( n+1\right) }{%
un\left( 2n+1\right) +3\left( k-p\right) \left( n+1\right) }\right\vert \label{C3}
\end{equation}

When $n\gg1$, the equation (\ref{C3}) can be approximated to

\begin{equation}
\Delta E_{n+1}^{m}=\frac{4\pi ^{2}\hbar ^{2}}{3ML^{2}}\left( n+1\right)
\left\vert \frac{3\left( k-2i\right) }{\frac{un\left( 2n+1\right) }{\left(
n+1\right) ^{2}}+\frac{3\left( k-p\right) }{\left( n+1\right) }}\right\vert
\approx \frac{4\pi ^{2}\hbar ^{2}}{ML^{2}}\left( n+1\right) \left\vert \frac{%
k}{2u}-\frac{p}{u}\right\vert \label{C4}
\end{equation}

{\bf Case (1)}: $0\leq k<2u$:

When $k$ is odd, in the large $n$ limit, $\overline{W_{0F}}$
is given by
\begin{subequations}
\label{C5whole}
\begin{equation}
\lim_{n\rightarrow \infty }\overline{W_{0F}}=\frac{8\pi ^{2}\hbar ^{2}}{%
ML^{2}}\lim_{n\rightarrow \infty }\frac{\sum_{p=0}^{\frac{k-1}{2}}p\frac{%
C_{2u}^{p}C_{2u}^{k-p}}{C_{4u}^{k}}\left( \frac{k}{2u}-\frac{p}{u}\right) }{%
\frac{4un+k}{n+1}}=\frac{\pi ^{2}\hbar ^{2}}{u^{2}ML^{2}}\sum_{p=0}^{\frac{%
k-1}{2}}p\frac{C_{2u}^{p}C_{2u}^{k-p}}{C_{4u}^{k}}\left( k-2p\right) \label{C5-1}
\end{equation}

Similarly, when $k$ is even, we have

\begin{equation}
\lim_{n\rightarrow \infty }\overline{W_{0F}}=\frac{\pi ^{2}\hbar ^{2}}{%
u^{2}ML^{2}}\sum_{p=0}^{\frac{k}{2}-1}p\frac{C_{2u}^{p}C_{2u}^{k-p}}{%
C_{4u}^{k}}\left( k-2p\right) \label{C5-2}
\end{equation}

{\bf Case (2)}: $2u\leq k<4u$

In the same way, when $k$ is odd, we have

\begin{equation}
\lim_{n\rightarrow \infty }\overline{W_{0F}}=\frac{\pi ^{2}\hbar ^{2}}{%
u^{2}ML^{2}}\sum_{q=0}^{\frac{4u-k-1}{2}}q\frac{C_{2u}^{q}C_{2u}^{4u-k-q}}{%
C_{4u}^{4u-k}}\left( 4u-k-2q\right) \label{C5-3}
\end{equation}

When $k$ is even, we get

\begin{equation}
\lim_{n\rightarrow \infty }\overline{W_{0F}}=\frac{\pi ^{2}\hbar ^{2}}{%
u^{2}ML^{2}}\sum_{q=0}^{\frac{4u-k}{2}-1}q\frac{C_{2u}^{q}C_{2u}^{4u-k-q}}{%
C_{4u}^{4u-k}}\left( 4u-k-2q\right) \label{C5-4}
\end{equation}
\end{subequations}
We can prove that $\overline{W_{0F}}$ becomes periodic and symmetric about $k=2u$ in the large $n$ limit.

\begin{acknowledgments}
The authors gratefully acknowledge the financial support of the project
from the Fundamental Research Fund for the Central Universities.
\end{acknowledgments}

\bibliographystyle{apsrev4-1}
\bibliography{QSZEref}

\end{document}